\documentclass[a4paper]{report}
\usepackage[utf8]{inputenc}
\usepackage[T1]{fontenc}
\usepackage{RJournal}
\usepackage{amsmath,amssymb,array}
\usepackage{booktabs}
\usepackage[english]{babel}
\usepackage{graphicx} 
\usepackage{multirow} 

\usepackage{caption}
\usepackage{subcaption}
\begin{document}

\sectionhead{Contributed research article}
\volume{Submitted 10/21/15}
\volnumber{YY}
\year{20ZZ}
\month{AAAA}

\begin{article}


\title{\pkg{QPot}: An R Package for Stochastic Differential Equation Quasi-Potential Analysis}
\author{by Christopher M. Moore, Christopher R. Stieha, Ben C. Nolting, Maria K. Cameron, and Karen C. Abbott}

\maketitle

\abstract{
\pkg{QPot} (pronounced $ky\overline{\textbf{o}o} + p\ddot{a}t$) is an R package for analyzing two-dimensional systems of stochastic differential
equations. It provides users with a wide range of tools to simulate, analyze, and visualize the dynamics of these systems. One of \pkg{QPot's} key features is the computation of the quasi-potential, an important tool for studying stochastic systems. Quasi-potentials are particularly useful for comparing the relative stabilities of equilibria in systems with alternative stable states. This paper describes \pkg{QPot}'s primary functions, and explains how quasi-potentials can yield insights about the dynamics of stochastic systems. Three worked examples guide users through the application of \pkg{QPot}'s functions.
}

\section*{Introduction}
Differential equations are an important modeling tool in virtually every scientific discipline. Most differential equation models are deterministic, meaning that they provide a set of rules for how variables change over time, and no randomness comes into play. Reality, of course, is filled with random events (i.e., noise or stochasticity). Unfortunately, many of the analytic techniques developed for deterministic ordinary differential equations are insufficient to study stochastic systems, where phenomena like noise-induced transitions between alternative stable states and metastability can occur.  For systems subject to stochasticity, the quasi-potential is a tool that yields information about properties such as the expected time to escape a basin of attraction, the expected frequency of transitions between basins, and the stationary probability distribution. \CRANpkg{QPot} (abbreviation of \strong{Q}uasi-\strong{Pot}ential) is an R package that allows users to calculate quasi-potentials, and this paper is a guided tutorial of its application.

\section*{Key functions}
\begin{tabular}{l p{4cm} p{7cm}}
\toprule
Function & Main arguments & Description \\
\midrule
\code{TSTraj()} & Deterministic skeleton, $\sigma$, $T$, $\Delta t$ & Creates a realization (time series) of the stochastic differential equations.\\
\code{TSPlot()} & \code{TSTraj()} output & Plots a realization of the stochastic differential equations, with an optional histogram side-plot.  Plots can additionally be two-dimensional, which show realizations in $(X,\,Y)$-space.\\
\code{TSDensity()} & \code{TSTraj()} output & Creates a density plot of a trajectory in $(X,\,Y)$-space in one or two dimensions.\\
\code{QPotential()} & Deterministic skeleton, stable equilibria, bounds, mesh (number of divisions along each axis) & Creates a matrix corresponding to a discretized version of the local quasi-potential function for each equilibrium.\\
\code{QPGlobal()} & Local quasi-potential matrices, unstable equilibria & Creates a global quasi-potential surface.\\ 
\code{QPInterp()} & Global quasi-potential, $(x\,y)$-coordinates & Evaluates the global quasi-potential at $(x,\,y)$.\\
\code{QPContour()} & Global quasi-potential & Creates a contour plot of the quasi-potential.\\
\code{VecDecomAll()} & Global quasi-potential, deterministic skeleton, bounds & Creates three vector fields: the deterministic skeleton, the negative gradient of the quasi-potential, and the remainder vector field. To find each field individually, the functions \code{VecDecomVec()}, \code{VecDecomGrad()}, or \code{VecDecomRem()} can be used.\\
\code{VecDecomPlot()} & Deterministic skeleton, gradient, or remainder field & Creates a vector field plot for the vector, gradient, or remainder field.\\
\bottomrule
\end{tabular}

\section*{Adding stochasticity to deterministic models}
Consider a differential equation model of the form
\begin{flalign}
\label{Deterministic}
\frac{dx}{dt}=f_{1}(x,y) \nonumber \\
\frac{dy}{dt}=f_{2}(x,y).
\end{flalign}
In many cases, state variables are subject to continual random perturbations, which are commonly modeled as white
noise processes. To incorporate these random influences, the original
system of deterministic differential equations can be transformed
into a system of stochastic differential equations:
\begin{flalign}
\label{Stochastic}
dX=f_{1}(X,Y)\, dt+\sigma\, dW_{1} \nonumber \\ 
dY=f_{2}(X,Y)\, dt+\sigma\, dW_{2}.
\end{flalign}
$X$ and $Y$ are now stochastic processes (a change emphasized through
the use of capitalization); this means that, at every time $t$, $X(t)$
and $Y(t)$ are random variables, as opposed to real numbers. $\sigma\geq0$ is a parameter specifying the noise intensity, and $W_{1}$ and $W_{2}$
are Wiener processes. A Wiener process is a special type of continuous-time stochastic process whose changes over non-overlapping time intervals, $\Delta t_{1}$ and $\Delta t_{2}$, are independent Gaussian random variables with mean zero and variances $\sqrt{\Delta t_{1}}$ and $\sqrt{\Delta t_{2}}$, respectively. The differential notation in equations~\eqref{Stochastic} is a formal way of representing a set of stochastic integral equations, which must be used because realizations of Wiener processes are not differentiable (to be precise, with probability one, a realization of a Wiener process will be almost nowhere differentiable). The functions $f_{1}$ and $f_{2}$ are called the deterministic skeleton. The deterministic skeleton can be viewed as a vector field that determines the dynamics of trajectories in the absence of stochastic effects.
We will forgo a complete overview of stochastic differential equations
here; interested readers are encouraged to seek out texts like \cite{Allen:2007ww} and \cite{iacus:2009aa}. We note, however, that throughout this paper
we use the It\^{o} formulation of stochastic differential equations.

\section*{The quasi-potential}
Consider system~\eqref{Stochastic}, with deterministic skeleton~\eqref{Deterministic}. If there exists
a function $V(x,y)$ such that $f_{1}(x,y)=-\frac{\partial V}{\partial x}$
and $f_{2}(x,y)=-\frac{\partial V}{\partial y}$, then system~\eqref{Deterministic} is called
a gradient system and $V(x,y)$ is called the system's potential function. The dynamics of a gradient system can be visualized by considering
the $\left(x,y\right)$-coordinates of a ball rolling on a surface
specified by $z=V(x,y)$. Gravity causes the ball to roll downhill,
and stable equilibria correspond to the bottoms of the surface's valleys.
$V(x,y)$ is a Lyapunov function for the system, which means that
if $\left(x(t),\, y(t)\right)$ is a solution to the system of equations~\eqref{Deterministic}, 
then $\frac{d}{dt}\left(V\left(x(t),\, y(t)\right)\right)\leq0$, 
and the only places that $\frac{d}{dt}\left(V\left(x(t),\, y(t)\right)\right)=0$ are at equilibria. This means that the ball's elevation will
monotonically decrease, and will only be constant if the ball is at
an equilibrium. The basin of attraction of a stable equilibrium $\mathbf{e}^{*}$
of system~\eqref{Deterministic} is the set of points that lie on solutions that asymptotically
approach $\mathbf{e}^{*}$.

The potential function is useful for understanding the stochastic
system~\eqref{Stochastic}. As in the deterministic case, the dynamics of the stochastic
system can be represented by a ball rolling on the surface $z=V(x,y)$;
in the stochastic system, however, the ball experiences random perturbations due to the noise terms in system~\eqref{Stochastic}. In systems with multiple stable equilibria,
these random perturbations can cause a trajectory to move between
different basins of attraction. The depth of the potential (that is,
the difference in $V$ at the equilibrium and the lowest point on
the boundary of its basin of attraction), is a useful measure of the
stability of the equilibrium \citep[see][]{nolting2015}. The deeper
the potential, the less likely it will be for stochastic perturbations
to cause an escape from the basin of attraction. This relationship
between the potential and the expected time to escape from a basin
of attraction can be made precise \citep[formulae in the appendices of][]{nolting2015}. Similarly, the potential function
is directly related to the expected frequency of transitions between
different basins, and to the stationary probability distribution of system~\eqref{Stochastic}.

Unfortunately, gradient systems are very special, and a generic system
of the form~\eqref{Deterministic} will almost certainly not be a gradient system. That
is, there will be no function $V(x,y)$ that satisfies $f_{1}(x,y)=-\frac{\partial V}{\partial x}$
and $f_{2}(x,y)=-\frac{\partial V}{\partial y}$ . Fortunately, quasi-potential
functions generalize the concept of a potential function for use in
non-gradient systems. The quasi-potential, $\Phi(x,y)$, of a non-gradient
system will possess many of the important properties that potential
functions have in gradient systems. The surface $z=\Phi(x,y)$ describes
the system's dynamics, and the depth of the surface is a highly useful
stability metric. Analogous to the potential described above, $\Phi(x,y)$ is directly related to the the stationary probability distribution, the expected frequency of transitions between basins of attraction, and the expected time required to escape each basin.

In both this paper and in \pkg{QPot}, the function that we refer to as the
quasi-potential is $\frac{1}{2}$ times the quasi-potential as defined
by Freidlin and Wentzell \citep{Freidlin:2012wd}. This choice is made so that the quasi-potential will agree with the potential in gradient systems.

\pkg{QPot} is an R package that contains tools for calculating and analyzing
quasi-potentials. The following three examples show how to use the tools
in this package. The first example is a simple consumer-resource model from ecology. This example is explained in detail, starting with the analysis of the deterministic skeleton, proceeding with simulation of the stochastic system, and finally demonstrating the calculation, analysis, and interpretation of the quasi-potential. The second and third examples are covered in less detail, but illustrate some special system behaviors. Systems with limit cycles, like example 2, require a slightly different procedure than systems that only have point attractors. Extra care must be taken constructing global quasi-potentials for exotic systems, like example 3. For more information about quasi-potentials, see \citet{Cameron:2012ex}, \citet{nolting2015}, and the references therein.

\section*{Example 1: A consumer-resource model with alternative stable states}
Consider the stochastic version (\textit{sensu} \eqref{Stochastic}) of a standard consumer-resource model of plankton ($X$) and their consumers ($Y$) \citep{collie:1994aa,steele:1981aa}:
\begin{flalign}
\label{ex1}
dX=\left(\alpha X\left(1-\frac{X}{\beta}\right)-\frac{\delta\, X^{2}\, Y}{\kappa+X^{2}}\right)\, dt+\sigma\, dW_{1} \nonumber \\
dY=\left(\frac{\gamma\, X^{2}\, Y}{\kappa+X^{2}}-\mu\, Y^{2}\right)\, dt+\sigma\, dW_{2}.
\end{flalign}
The model is formulated with a Type III functional response, meaning that the highest per-capita consumption rate of plankton occurs at intermediate plankton densities. $\alpha$ is the plankton's maximum population growth rate, $\beta$ is the plankton carrying capacity, $\delta$ is the maximal feeding rate of the consumers, $\gamma$ is the conversion rate of plankton to consumer, and $\mu$ is the consumer mortality rate. We will analyze this example with a set of parameter values that yield two stables states: $\alpha=1.54$, $\beta=10.14$, $\gamma=0.476$, $\delta=1$, $\kappa=1$, and $\mu=0.112509$.

\subsection*{Step 1: Analyzing the deterministic skeleton}
There are preexisting tools in R for analyzing the deterministic skeleton of system \eqref{ex1}, which will be described briefly in this subsection. The first step is to find the equilibria for the system and determine their stability with linear stability analysis. Equilibria can be found using the package \CRANpkg{rootSolve} \citep{soetaert:2008aa}.  In example 1, the equilibria are $\mathbf{e}_{u1}=(0,0)$, $\mathbf{e}_{s1}=(1.4049,\,2.8081)$,
$\mathbf{e}_{u2}=(4.2008,\,4.0039)$, $\mathbf{e}_{s2}=(4.9040,\,4.0619)$, and $\mathbf{e}_{u3}=(10.14,\,0)$. The package \CRANpkg{deSolve} \citep{soetaert:2010aa} can find the eigenvalues of the linearized system at an equilibrium, which determines the asymptotic stability of the system. $\mathbf{e}_{u1}$ is an unstable source and $\mathbf{e}_{u2}$ and $\mathbf{e}_{u3}$ are saddles. The eigenvalues corresponding to $\mathbf{e}_{s1}$ are $-0.047\pm0.548\, i$ and the eigenvalues corresponding
to $\mathbf{e}_{s2}$ are $-0.377$ and $-0.093$. Hence $\mathbf{e}_{s1}$ is a stable spiral point and $\mathbf{e}_{s2}$ is a stable node. To ease transition from packages such as \pkg{deSolve} to our package \pkg{QPot}, we include the wrapper function \code{Model2String()}, which takes a function containing equations and a list of parameters and their values, and returns the equations in a string that is usable by \pkg{QPot}.

\begin{figure}[ht]
  \centering
  \includegraphics[width=300px]{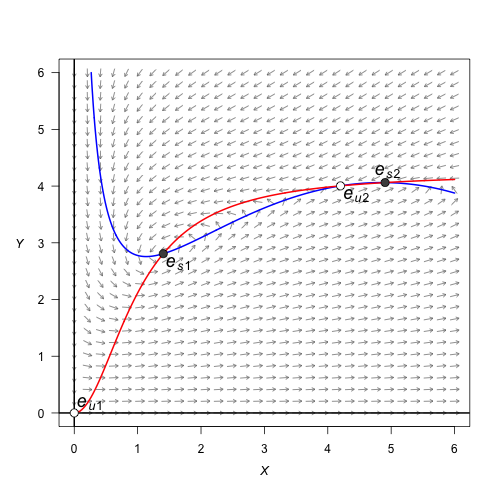}
  \caption{A stream plot of the deterministic skeleton of system \eqref{ex1}. The blue line is an $x$-nullcline (where $\frac{dx}{dt}=0$) and the red line is a $y$-nullcline (where $\frac{dy}{dt}=0$). Open circles are unstable equilibria and filled circles are stable equilibria. Made using the package \pkg{phaseR}.}
  \label{fig:ex1stream}
\end{figure}

The package \CRANpkg{phaseR} \citep{grayling:2014aa} generates a stream plot of the deterministic skeleton of the system of equations~\eqref{ex1} (Figure~\ref{fig:ex1stream}). Further, \pkg{deSolve} can be used to find solutions corresponding to particular initial conditions of the deterministic skeleton of system~\eqref{ex1}.  During the analysis of the deterministic skeleton of a system, it is important to note several things. The first is the range of $x$ and $y$ values over which relevant dynamics occur. In example 1, transitions between the stable equilibria are a primary point of interest, so one might wish to focus on a region like the one displayed in Figure~\ref{fig:ex1stream}, even though this region excludes $\mathbf{e}_{u3}$. The ranges of the variables will determine the window sizes and ranges used later in the quasi-potential calculations. Second, it is important to note if there are any limit cycles. If there are, it will be necessary to identify a point on the limit cycle. This can be accomplished by calculating a long-time solution of the system of ODEs to obtain a trajectory that settles down on the limit cycle (see example 2).  Finally, it is important to note regions of phase space that correspond to unbounded solutions. As explained in subsequent sections, it is worth examining system behavior in negative phase space, even in cases where negative quantities lack physical meaning.

\subsection*{Step 2: Stochastic simulation}
For a specified level of noise intensity, $\sigma,$ one can obtain a realization of system~\eqref{ex1}. To do this, \code{TSTraj()} in \pkg{QPot} implements the Euler-Maruyama method. All other code/function references hereafter are found in \pkg{QPot}, unless specified otherwise. To generate a realization, the following arguments are required: the right-hand side of the deterministic skeleton for both equations, the initial conditions $\left(x_{0},\,y_{0}\right)$, the parameter values, the step-size $\Delta t$, and the total time length $T$. 

\begin{example}
var.eqn.x <- "(alpha*x)*(1-(x/beta)) - ((delta*(x^2)*y)/(kappa+(x^2)))"
var.eqn.y <- "((gamma*(x^2)*y)/(kappa+(x^2))) - mu*(y^2)"
model.state <- c(x = 1, y = 2)
model.parms <- c(alpha = 1.54, beta = 10.14, delta = 1, gamma = 0.476, 
   kappa = 1, mu = 0.112509)
model.sigma <- 0.05
model.time <- 1000 # we used 12500 in the figures
model.deltat <- 0.025
ts.ex1 <- TSTraj(y0 = model.state, time = model.time, deltat = model.deltat, 
     x.rhs = var.eqn.x, y.rhs = var.eqn.y, parms = model.parms, sigma = model.sigma)
\end{example}

\begin{figure}[ht]
\begin{subfigure}[b]{0.4\textwidth}
\caption{}
\includegraphics[width=400px]{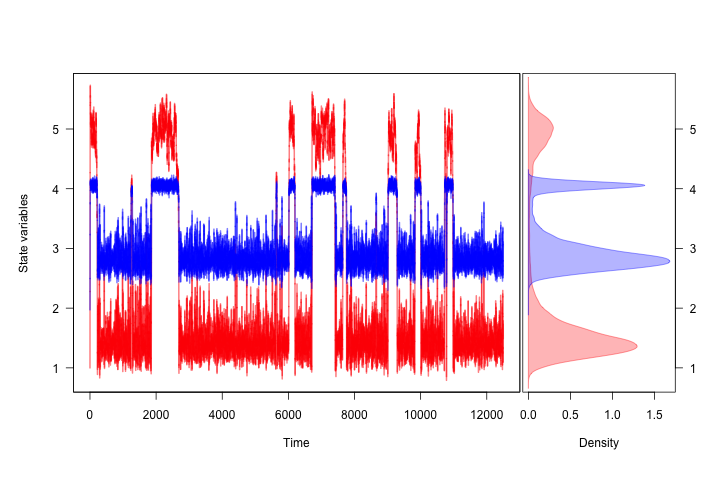}
\end{subfigure}\\
\begin{subfigure}[b]{0.4\textwidth}
\caption{}
\includegraphics[width=300px]{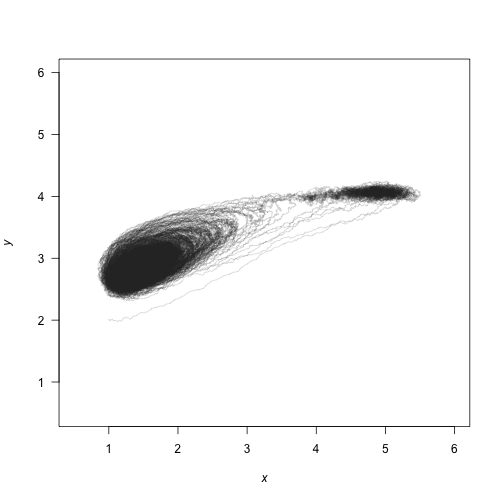}
\end{subfigure}
\caption{\textbf{(a)} A realization of system~\eqref{ex1} created using \code{TSPlot()}, with $x$ in blue and $y$ in red. The left side of (a) shows the time series. The right side of (a), which is enabled with the default \code{dens = T}, shows a histogram of the $x$ and $y$ values over the entire realization. \textbf{(b)} The realization plotted in $(x,y)$-space with \code{dim = 2}.}
\label{fig:ex1ts}
\end{figure}

\begin{figure}[ht]
\centering
\includegraphics[width=300px]{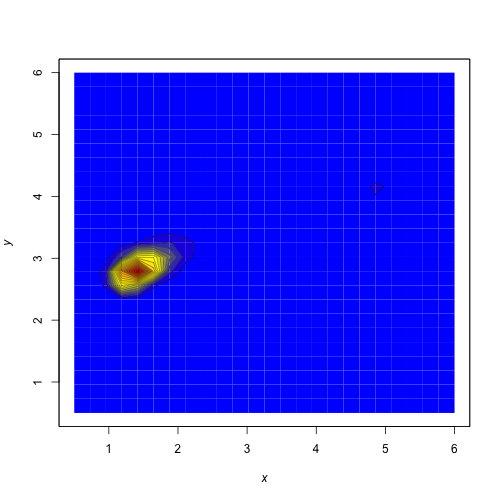}
\caption{A density plot obtained from a realization of system~\eqref{ex1}. Red corresponds to high density, and blue to low density.  Plotted using the function \code{TSDensity()} with \code{dim = 2}.}
\label{fig:ex1dens}
\end{figure}

Figure~\ref{fig:ex1ts} shows a realization for $\sigma=0.05,$  $\Delta t=0.025,$ $T=1.25\times10^{4},$  and initial condition $(x_{0},\, y_{0})=(1,\,2).$  The argument \code{dim = 1} produces a time series plot with optional histogram side-plot. The \code{dim = 2} produces a plot of a  realization in $(x,y)$-space. If the system is ergodic, a very long realization will approximate the steady-state probability distribution. Motivated by this, a probability density function can be approximated from a long realization using the \code{TSDensity()} function (e.g., Figure~\ref{fig:ex1dens}).

\begin{example}[commandchars=\\\{\}]
TSPlot(ts.ex1, deltat = model.deltat)		# Figure \ref{fig:ex1ts}a
TSPlot(ts.ex1, deltat = model.deltat, dim = 2)	# Figure \ref{fig:ex1ts}a
TSDensity(ts.ex1, dim = 1)		# like Figure \ref{fig:ex1ts}a histogram
TSDensity(ts.ex1, dim = 2)			# Figure \ref{fig:ex1dens}
\end{example}

Bounds can be placed on the state variables in all of the functions described in this subsection. For example, it might be desirable to set 0 as the minimum size of a biological population, because negative population densities are not physically meaningful. A lower bound can be imposed on the functions described in this subsection with the argument \code{lower.bound} in the function \code{TRTraj()}. Similarly, it might be desirable to set an upper bound for realizations, and hence prevent runaway trajectories (unbounded population densities are also not physically meaningful). An upper bound can be imposed on the functions described in this subsection with the argument \code{upper.bound}.

\subsection*{Step 3: Local quasi-potential calculation}
The next step is to compute a local quasi-potential for each attractor. Because \pkg{QPot} deals with two-dimensional systems, ``attractor" will be used synonymously with ``stable equilibrium" ``or stable limit cycle". A limit cycle will be
considered in example 2. For now, suppose that the only attractors
are stable equilibrium points, $\mathbf{e}_{si}$, $i=1,\,\ldots,\, n.$
In the example above, $n=2.$ For each stable equilibrium $\mathbf{e}_{si}$,
we will compute a local quasi-potential $\Phi_{i}(x,y).$

In order to understand the local quasi-potential, it is useful consider the analogy of a particle traveling according to system~\eqref{Stochastic}. In the context of example 1, the coordinates of the particle correspond to population densities, and the particle's path corresponds to how those population densities change over time. The deterministic skeleton of~\eqref{Stochastic} can be visualized as a force field influencing the particle's trajectory. Suppose that the particle
moves along a path from a stable equilibrium $\mathbf{e}_{si}$
to a point $(x,y)$. If this path does not coincide with a solution
of the deterministic skeleton, then the stochastic terms must be doing
some ``work'' to move the particle along the path. The more work
that is required, the less likely it is for the path to be a realization
of system~\eqref{Stochastic}. $\Phi_{i}(x,y)$ is the amount of work required to traverse
the easiest path from $\mathbf{e}_{si}$ to $(x,y)$. Note that $\Phi_{i}(x,y)$
is non-negative, and it is zero at $\mathbf{e}_{si}$.

In the basin of attraction for $\mathbf{e}_{si}$, $\Phi_{i}(x,y)$
has many properties analogous to the potential function for gradient
systems. Key among these properties is that the quasi-potential is non-increasing
along deterministic trajectories. This means that the quasi-potential
can be interpreted as a type of energy surface, and the rolling ball
metaphor is still valid. The difference is that, in non-gradient systems,
there is an additional component to the vector field that causes trajectories
to circulate around level sets of the energy surface. This is discussed
in more detail in Step 6, below.

\pkg{QPot} calculates quasi-potentials using an adjustment developed by \citet{Cameron:2012ex} to the ordered upwind algorithm \citep{sethian:2001aa,sethian:2003aa}. The idea behind the algorithm is
to calculate $\Phi_{i}(x,y)$ in ascending order, starting with the
known point $\mathbf{e}_{si}$. The result is an expanding area where
the solution is known.

Calculating $\Phi_{i}(x,y)$ with the function \code{QPotential()} requires a text string of the equations and parameter values, the stable equilibrium points, the computation domain, and the mesh size. For~\eqref{ex1}, this first means inputting the equations:
\begin{flalign*}
f_{1}(x,y)=1.54x\left(1-\frac{x}{10.14}\right)-\frac{(1)x^{2}\, y}{1+x^{2}}\\
f_{2}(x,y)=\frac{0.476\, x^{2}\, y}{1+x^{2}}-0.112509\, y^{2}.
\end{flalign*}

\noindent In R:

\begin{example}
equation.x = "1.54*x*(1.0-(x/10.14))-(y*x*x)/(1.0+x*x)"
equation.y = "((0.476*x*x*y)/(1+x*x))-0.112590*y*y"
\end{example}

The coordinates of the points $\mathbf{e}_{si}$, which were determined in Step 1, are $\mathbf{e}_{s1}=(1.4049,\,2.8081)$ and $\mathbf{e}_{s2}=(4.9040,\,4.0619)$. 

\begin{example}
eq1.x = 1.40491
eq1.y = 2.80808
eq2.x = 4.9040
eq2.y = 4.06187
\end{example}

Next, the boundaries of the computational domain need to be entered. This domain will be denoted by 
$\left[Lx_{1},Lx_{2}\right]\times\left[Ly_{1},Ly_{2}\right]$. The ordered-upwind method terminates when the solved area encounters
a boundary of this domain. Thus, it is important to choose boundaries
carefully. For example, if $\mathbf{e}_{si}$ lies on one of the coordinate
axes, one should not use that axis as a boundary because the algorithm will immediately terminate. Instead, one should add padding space.
This is important even if the padding space corresponds to physically
unrealistic values (e.g., negative population densities). For this
example, a good choice of boundaries is: $Lx_{1}=Ly_{1}=-0.5,$ and $Lx_{2}=Ly_{2}=20$. This choice of domain was obtained by examining stream plots of the deterministic skeleton and density plots of stochastic realizations (Figures \ref{fig:ex1stream}--\ref{fig:ex1dens}). The domain contains all of the deterministic skeleton equilibria, and it encompasses a large area around the regions of phase space visited by stochastic trajectories (Figures \ref{fig:ex1stream}--\ref{fig:ex1dens}). Note that a small padding space was added to the left and bottom sides of the domain, so that the coordinate axes are not the domain boundaries.

\begin{example}
bounds.x = c(-0.5, 20.0)
bounds.y = c(-0.5, 20.0)
\end{example}

In some cases, it may be desirable to treat boundaries differently
in the upwind algorithm. This is addressed below in the section ``Boundary behavior''.

Finally, the mesh size for the discretization of the domain needs to be specified. 
Let $N_{x}$ be the number of grid points in the $x$-direction and $N_{y}$ be the number of grid points in the $y$-direction. Note that the horizontal distance between mesh points is $h_{x}=\frac{Lx_{2}-Lx_{1}}{N_{x}}$,
and the vertical distance between mesh points is $h_{y}=\frac{Ly_{2}-Ly_{1}}{N_{y}}.$
Mesh points are considered adjacent if their Euclidean distance is
less than or equal to $h=\sqrt{h_{x}^{2}+h_{y}^{2}}.$ This means
that diagonal mesh points are considered adjacent. In this example,
a good choice is $N_{x}=N_{y}=4100.$ This means that $h_{x}=h_{y}=0.005,$ and
$h\approx0.00707.$ In general, the best choice of mesh size will be a compromise between resolution and computational time. The mesh size must be fine enough to precisely track how information moves outward along characteristics from the initial point. Too fine of a mesh size can lead to very long computational times, though. The way that computation time scales with grid size depends on the system under consideration (see below for computation time for this example), because the algorithm ends when it reaches a boundary, which could occur before the algorithm has exhaustively searched the entire mesh area.

\begin{example}
step.number.x = 1000 # we used 4100 in the figures
step.number.y = 1000 # we used 4100 in the figures
\end{example}

The ``anisotropy ratio'' is another adjustable parameter for the
algorithm, defined by \code{k.x} and \code{k.y} in \code{QPotential()}.  For more on this, see \cite{Cameron:2012ex}. For now, we suggest
using the defaults $K_{x}=20$ and $K_{y}=20$.

The R interface implements the \code{QPotential()} algorithm using C
code. By default \code{QPotential()} outputs a matrix that contains the quasi-potentials to the R session. 
The time required to compute the quasi-potential will depend on the size of the region and the fineness of the mesh. 
This example with $K_{x}=K_{y}=20$ and $N_{x}=N{y}=4100$ has approximately $1.7\times10^{7}$
grid points, which leads to run times of approximately 2.25 min (2.5 GHz Intel Core i5 processor and 8 GB 1600 MHz DDR3 memory). 
When one reaches around $5\times10^{8}$, computational time can be several hours. Setting the argument \code{save.to.R} to \code{TRUE} outputs the matrix into the R session, and setting the argument \code{save.to.HD} to \code{TRUE} saves the matrix to the hard drive as the file \code{filename} in the current working directory. For $N_{x}=N{y}=4100$, the saved file occupies 185 MB.

\begin{example}
eq1.local <- QPotential(x.rhs = equation.x, x.start = eq1.x, x.bound = bounds.x, 
   x.num.steps = step.number.x, y.rhs = equation.y, y.start = eq1.y,  y.bound = 
   bounds.y, y.num.steps = step.number.y)
\end{example}

Step 3 should be repeated until local quasi-potentials $\Phi_{i}(x,y)$ have been obtained
for each $\mathbf{e}_{si}$. In example 1, this means calculating
$\Phi_{1}(x,y)$ corresponding to $\mathbf{e}_{s1}$ and $\Phi_{2}(x,y)$
corresponding to $\mathbf{e}_{s2}$.

\begin{example}
eq2.local <- QPotential(x.rhs = equation.x, x.start = eq2.x, x.bound = bounds.x, 
   x.num.steps = step.number.x, y.rhs = equation.y, y.start = eq2.y, y.bound = 
   bounds.y, y.num.steps = step.number.y)
\end{example}

Each local quasi-potential $\Phi_{i}(x,y)$ is stored in R as a large matrix. The entries in this matrix are the values of $\Phi_{i}$ at each mesh point. To define the function on the entire domain (i.e., to allow it to be evaluated at arbitrary points in the domain, not just the discrete mesh points), bilinear interpolation is used. The values of $\Phi(x,y)$ can be extracted using the function \code{QPInterp()}. Inputs to \code{QPInterp()} include the ($x,\,y$) coordinates of interest, the ($x,\,y$) domain boundaries, and the \code{QPotential()} output (i.e., the matrix with rows corresponding to $x$-values and columns corresponding to $y$-values).  \code{QPInterp()} can be used for any of the local quasi-potential or the global quasi-potential surfaces (see the next subsection).

\subsection*{Step 4: Global quasi-potential calculation}
Recall that $\Phi_{i}(x,y)$ is the amount of ``work'' required
to travel from $\mathbf{e}_{si}$ to $(x,y)$. This information is
useful for considering dynamics in the basin of attraction of $\mathbf{e}_{si}$.
In many cases, however, it is desirable to define a global quasi-potential
that describes the system's dynamics over multiple basins of attraction.
If a gradient system has multiple stable states, the potential function
provides an energy surface description that is globally valid. We
seek an analogous global function for non-gradient systems. Achieving
this requires ``pasting'' local quasi-potentials into a single
global quasi-potential. If the system has only two 
attractors, one can define a global quasi-potential, though, it might be nontrivial, see example 3 ahead.
In systems with three or more attractors such a task might not be possible 
 \citep{Freidlin:2012wd}. For a wide variety of systems, however, a relatively simple algorithm can accomplish
the pasting \citep{Graham:1986cq, Roy:1995tx}. In most cases, the algorithm amounts to translating the local quasi-potentials up or down so that they agree at the saddle points that separate the basins of attraction. In example 1, $\mathbf{e}_{u1}$ lies on the boundary of the basins of attraction for $\mathbf{e}_{s1}$ and $\mathbf{e}_{s2}$. Creating a global quasi-potential requires matching $\Phi_{1}$ and $\Phi_{2}$ at $\mathbf{e}_{u2}.$ $\Phi_{1}(\mathbf{e}_{u2})=0.007056$ and $\Phi_{2}(\mathbf{e}_{u2})=0.00092975.$ If one defines 
\[ \Phi_{2}^{*}(x,y)=\Phi_{2}(x,y)+\left(0.007056-0.00092975\right)=\Phi_{2}(x,y)+0.00612625,\]
then $\Phi_{1}$ and $\Phi_{2}^{*}$ match at $\mathbf{e}_{u2}$. Finally, define
\[\Phi(x,y)=\min(\Phi_{1}(x,y),\Phi_{2}^{*}(x,y)), \]
which is the global quasi-potential. For systems with more than two stable equilibria, this process is generalized to match local quasi-potentials at appropriate saddles. \code{QPot} automates this procedure. A fuller description of the underlying algorithm is explained in example 3, which requires a more nuanced understanding of the pasting procedure.

\begin{example}
ex1.global <- QPGlobal(local.surfaces = list(eq1.local, eq2.local), 
   unstable.eq.x = c(0, 4.2008), unstable.eq.y = c(0, 4.0039), x.bound = bounds.x, 
   y.bound = bounds.y)
\end{example}

This function \code{QPGlobal} calculates the global quasi-potential by automatically pasting together the local quasi-potentials. This function requires the input of all the discretized local quasi-potentials, and the coordinates of all of the unstable equilibria. The output is a discretized version of the global quasi-potential. The length of time required for this computation will depend on the total number of mesh points; for the parameters used in example 1, it takes a couple of minutes. As with the local quasi-potentials, the values of $\Phi(x,y)$ can be extracted using the function \code{QPInterp()}.

\subsection*{Step 5: Global quasi-potential visualization}
\begin{figure}[ht]
  \centering
  \includegraphics[width=14cm]{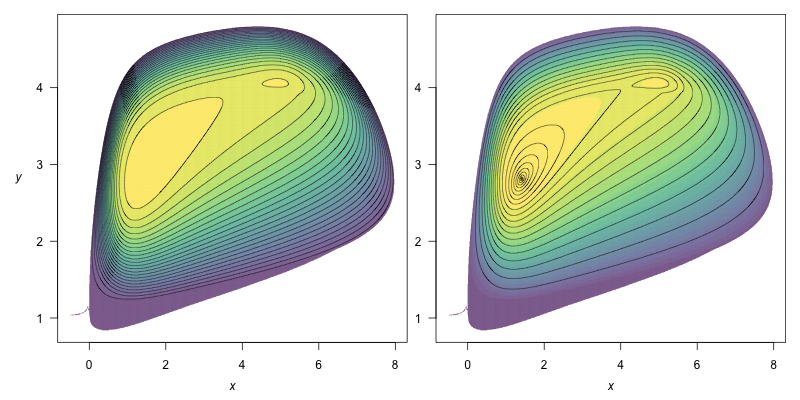}
  \caption{A contour plot of the of the quasi-potential of system~\eqref{ex1}. Yellow corresponds to low values of the quasi-potential, and purple to high values.  \code{c.parm} in \code{QPContour()}, can be used to condense the contour lines at the bottom of the basins for better resolution. The default creates evenly spaced contour lines (left; \code{c.parm = 1}). On the right, contour lines are condensed (\code{c.parm = 5}).}
  \label{fig:ex1qp}
\end{figure}

To visualize the global quasi-potential, one can simply take the global quasi-potential matrix from \code{QPGlobal} and use it to create a contour plot using \code{QPContour()} (Figure~\ref{fig:ex1qp}). 

\begin{example}
QPContour(surface = ex1.global, dens = c(1000, 1000), x.bound = bounds.x, 
   y.bound = bounds.y, c.parm = 5) # right side of Figure~\ref{fig:ex1qp}
\end{example}

\code{QPContour()} is based on the \code{.filled.contour()} function from the base package \pkg{graphics}. In most cases, the mesh sizes used for the quasi-potential calculation will be much finer than what is required for useful visualization. The argument \code{dens} within \code{QPContour()} reduces the points used in the graphics generation. Although it might seem wasteful to perform the original calculations at a mesh size that is finer than the final visualization, this is not so. Choosing the mesh size in the original calculations to be very fine reduces the propagation of errors in the ordered upwind algorithm, and hence leads to a more accurate numerical solution.

An additional option allows users to specify contour levels. R's default for the \code{contour()} function creates contour lines that are equally spaced over the range of values specified by the user. In some cases, however, it is desirable to use a non-linear spacing for the contours. For example, equally-spaced contours will not capture the topography at the bottom of a basin if the changes in height are much smaller than other regions in the plot. Simply increasing the number of equally-spaced contour lines does not solve this problem, because steep areas of the plot become completely saturated with lines. \code{QPContour()} has a function for non-linear contour spacing that condenses contour lines at the bottoms of basins. Specifically, for $n$ contour lines, this function generates a list of contour levels, $\left\{ v_{i}\right\} _{i=1}^{n}$, specified by:
\[ v_{i}= \max(\Phi) \left( \frac{i-1}{n -1} \right)^c. \]
$c = 1$ yields evenly-spaced contours. As $c$ increases, the contour lines become more concentrated near basin bottoms. Figure \ref{fig:ex1qp} shows equal contour lines (left panel) and contour lines that are concentrated at the bottom of the basin (right panel, \code{c.parm = 5}).

\begin{figure}[ht]
  \centering
  \includegraphics[width=14cm]{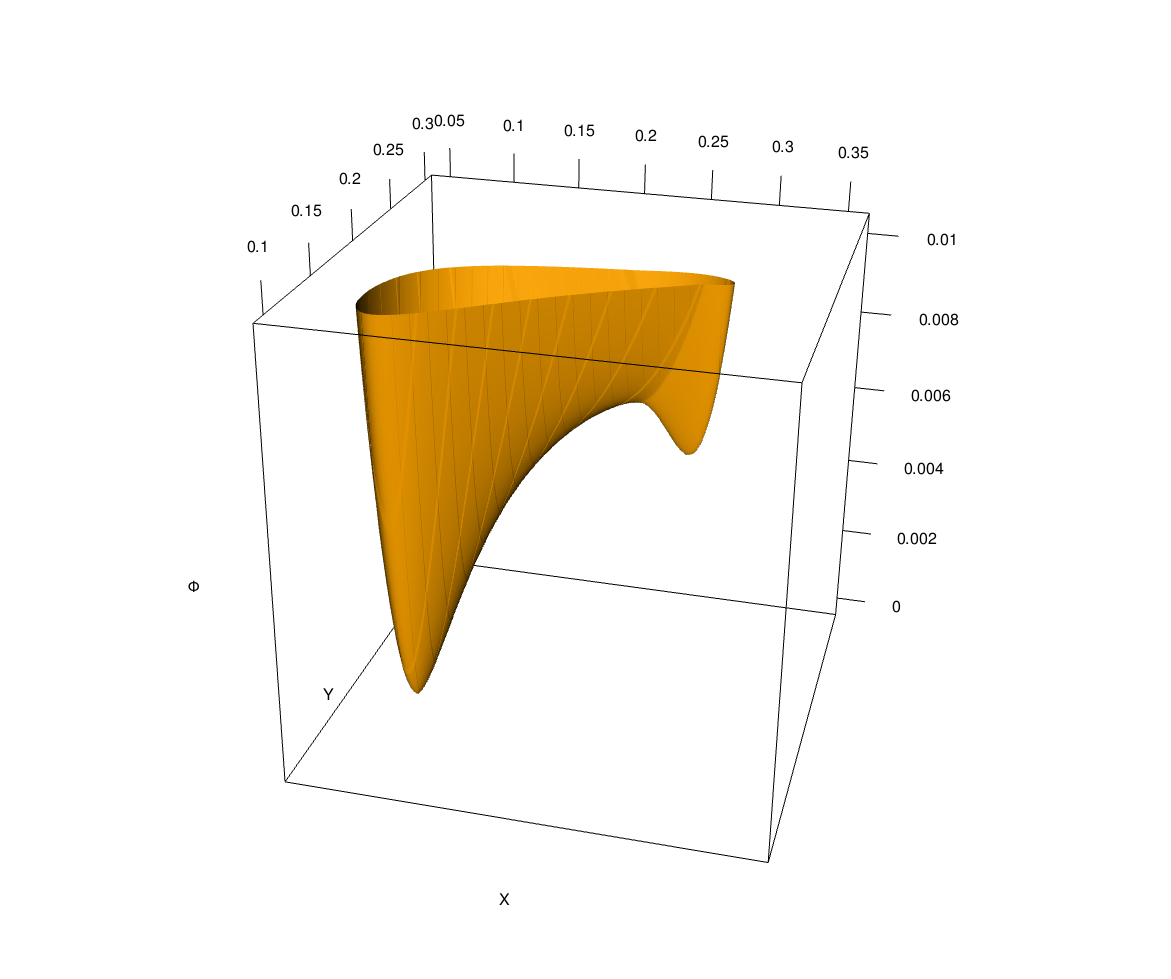}
  \caption{A 3D plot of the of the quasi-potential of system~\eqref{ex1} using \code{persp3d()} in package \pkg{rgl}.  3D plotting can further help users visualize the quasi-potential surfaces.}
  \label{fig:ex13d}
\end{figure}

Finally, creating a 3D plot can be very useful for visualizing the features of more complex surfaces. This is especially helpful when considering the physical metaphor of a ball rolling on a surface specified by a quasi-potential \citep{nolting2015}. R has several packages for 3D plotting, including static plotting with the base function \code{persp()} and with the package \CRANpkg{plot3D} \citep{soetaert:2013aa}. Interactive plotting is provided by \CRANpkg{rgl} \citep{adler:2015aa}. To create an interactive 3D plot for example 1 using \pkg{rgl}, use the code: \code{persp3d(x = ex1.global, col = "orange")}. Figure \ref{fig:ex13d} shows a 3D plot of example~1 that clearly illustrates the differences between the two local basins. Users can also export the matrix of quasi-potential values and create 3D plots in other programs.

\subsection*{Step 6: Vector field decomposition}
Recall that the deterministic skeleton~\eqref{Deterministic} can be visualized as a vector field,
as shown in Figure~\ref{fig:ex1stream}.  In gradient systems, this vector field is
completely determined by the potential function, $V(x,y)$. The name ``gradient system'' refers to the fact that the vector field
is the negative of the potential function's gradient,
\[ \renewcommand{\arraystretch}{1.4} \begin{bmatrix}f_{1}(x,y)\\ f_{2}(x,y) \end{bmatrix}=-\nabla V(x,y)= -\begin{bmatrix}\frac{\partial V}{\partial x}(x,y)\\ \frac{\partial V}{\partial y}(x,y) \end{bmatrix}. \]
In non-gradient systems, the vector field can no longer be represented
solely in terms of the gradient of $\Phi(x,y).$ Instead, there is
a remainder component of the vector field, $\mathbf{r}(x,y)=\begin{bmatrix}r_{1}(x,y)\\
r_{2}(x,y)
\end{bmatrix}.$ The vector field can be decomposed into two terms: 
\[ \renewcommand{\arraystretch}{1.4} \begin{bmatrix}f_{1}(x,y)\\ f_{2}(x,y) \end{bmatrix}=-\nabla \Phi(x,y)+\mathbf{r}(x,y)= -\begin{bmatrix}\frac{\partial \Phi}{\partial x}(x,y)\\ \frac{\partial \Phi}{\partial y}(x,y) \end{bmatrix} + \begin{bmatrix}r_{1}(x,y)\\ r_{2}(x,y) \end{bmatrix}.\]
The
remainder vector field is orthogonal to the gradient of the quasi-potential
everywhere. That is, for every $\left(x,y\right)$ in the domain,
\[
\nabla\Phi(x,y)\cdot\mathbf{r}(x,y)=0.
\]
An explanation of this property can be found in \citet{nolting2015}.

\begin{figure}[ht]
\centering
\includegraphics[width=14cm]{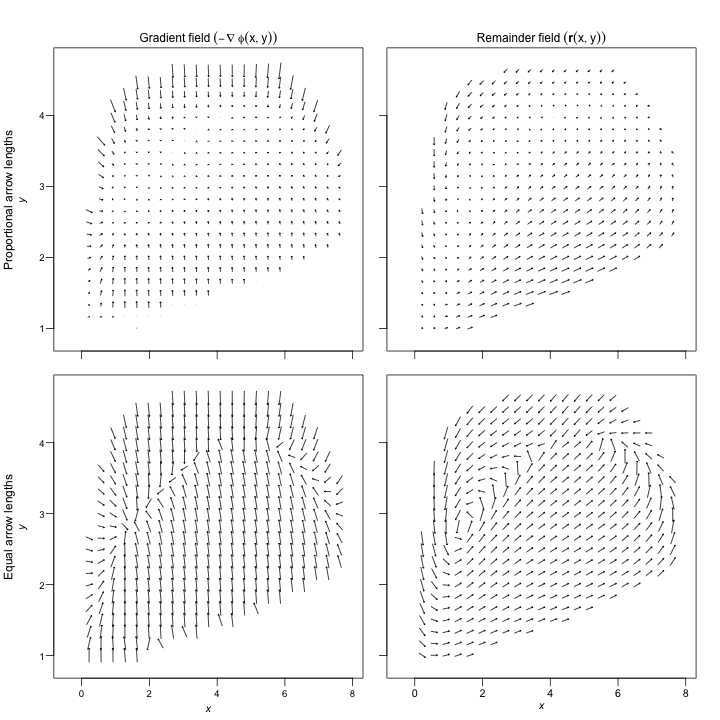}
\caption{The gradient (left column) and remainder (right column) fields, plotted with \code{arrow.type = "proportional"} (top row) and \code{arrow.type = "equal"} (bottom row) arrow lengths using \code{VecDecomPlot()} for system~\eqref{ex1}.}
\label{fig:ex1VF} 
\end{figure}

The remainder vector field can be interpreted as a force that causes trajectories to circulate around level sets of the quasi-potential. \pkg{QPot} enables users to perform this decomposition. The function \code{VecDecomAll()} calculates the vector field decomposition, and outputs three vector fields: the original deterministic skeleton, $\mathbf{f}(x,y)$; the gradient vector field, $-\nabla\Phi(x,y)$; and the remainder vector field, $\mathbf{r}(x,y)$. Each of these three vector fields can be output alone using \code{VecDecomDS()}, \code{VecDecomGrad()}, or \code{VecDecomRem()}. These vector fields can be visualized using the function \code{VecDecomPlot()}.  Code to create the vector fields from \code{VecDecomAll()} is displayed below; code for generating individual vector fields can be found in \code{help()} for \code{VecDecomDS()}, \code{VecDecomGrad()}, or \code{VecDecomRem()}.  The gradient and remainder vector fields are shown in the left and right columns of figure~\ref{fig:ex1VF}, respectively, with proportional vectors (top row) and equal-length vectors (bottom row).  Three arguments within \code{VecDecomPlot()} are important to creating comprehensible plots: \code{dens}, \code{tail.length}, and \code{head.length}. \code{dens} specifies the number of arrows in the plot window along the $x$ and $y$ axes. The argument \code{tail.length} scales the length of arrow tails. The argument \code{head.length} scales the length of arrow heads. The function \code{arrows()} makes up the base of \code{VecDecomPlot()}, and arguments can be passed to it, as well as to \code{plot}. The code below produces all three vector fields from the multi-dimensional array returned by \code{VecDecomAll()}:

\begin{example}
# Calculate all three vector fields
VDAll <- VecDecomAll(surface = ex1.global, x.rhs = equation.x, y.rhs = 
     equation.y, x.bound = bounds.x, y.bound = bounds.y)
# Plot the deterministic skeleton vector field
VecDecomPlot(field = list(VDAll[,,1], VDAll[,,2]), dens = c(25, 25), 
     x.bound = bounds.x, y.bound = bounds.y, x.lim = c(0, 11), y.lim = c(0, 6), 
     arrow.type = "equal", tail.length = 0.25, head.length = 0.025)
# Plot the gradient vector field
VecDecomPlot(field = list(VDAll[,,3], VDAll[,,4]), dens = c(25, 25), 
     x.bound = bounds.x, y.bound = bounds.y, arrow.type = "proportional", 
     tail.length = 0.25, head.length = 0.025)
# Plot the remainder vector field
VecDecomPlot(field = list(VDAll[,,5], VDAll[,,6]), dens = c(25, 25), 
     x.bound = bounds.x, y.bound = bounds.y, arrow.type = "proportional", 
     tail.length = 0.35, head.length = 0.025)
\end{example}

\section*{Example 2: A model with a limit cycle}
Consider the following model:
\begin{flalign}
\label{ex2}
dX=\left(-(Y-\beta )+\mu\,(X-\alpha)\left(1-(X-\alpha)^{2}-(Y-\beta )^{2}\right)\right)\, dt+\sigma\, dW_{1} \nonumber \\
dY=\left((X-\alpha)+\mu\,(Y-\beta )\left(1-(X-\alpha)^{2}-(Y-\beta )^{2}\right)\right)\, dt+\sigma\, dW_{2}.
\end{flalign}
We will analyze this example with $\mu=0.2$, $\alpha=4$, and $\beta=5$.

\subsection*{Step 1: Analyzing the deterministic skeleton}
\begin{figure}[ht]
  \centering
  \includegraphics[width=300px]{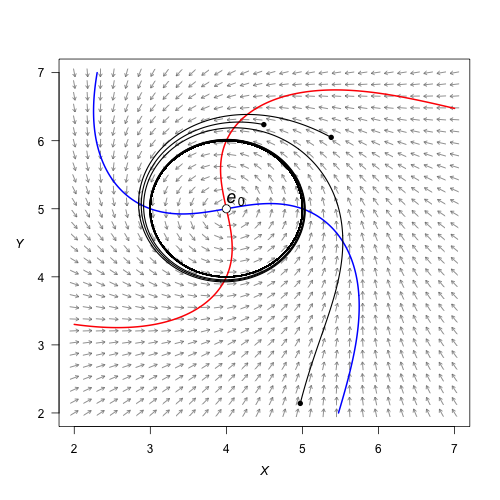}
  \caption{A stream plot of the deterministic skeleton of system~\eqref{ex2}. The blue line is an $x$-nullcline (where $\frac{dx}{dt}=0$) and the red line is a $y$-nullcline (where $\frac{dy}{dt}=0$). The open circle is an unstable equilibrium. Particular solutions are shown as black lines, with filled circles as initial conditions. Made using the package \pkg{phaseR}.}
  \label{fig:ex2stream}
 \end{figure}

The deterministic skeleton of this system has one equilibrium, $\mathbf{e}_{0}=(4,\,5),$
which is an unstable spiral point. Figure~\ref{fig:ex2stream} shows a stream
plot of the deterministic skeleton of system~\eqref{ex2}. A particular
solution of the deterministic skeleton of system~\eqref{ex2} can be found using \pkg{rootSolve} and \pkg{deSolve}.
The stream plot and a few particular solutions suggest that there
is a stable limit cycle. To calculate the limit cycle, once can find a particular
solution over a long time interval (e.g., Figure~\ref{fig:ex2stream} has three trajectories run for $T = 100$). The solution will eventually converge to the limit cycle. One can drop the early part
of the trajectory until only the closed loop of the limit cycle remains. There are more elegant ways to numerically find a periodic orbit (even when those orbits are unstable). For more information on these methods, see \citet{chua}.
In this example, the limit cycle is shown by the thick black line in Figure~\ref{fig:ex2stream}. For calculation of the quasi-potential,
it is sufficient to input a single point that lies on the limit
cycle. For this example, one such point is $\mathbf{z}=(4.15611,\,5.98774).$

\subsection*{Step 2: Stochastic simulation}
\begin{figure}[ht]
\begin{subfigure}[b]{0.4\textwidth}
\caption{}
\includegraphics[width=400px]{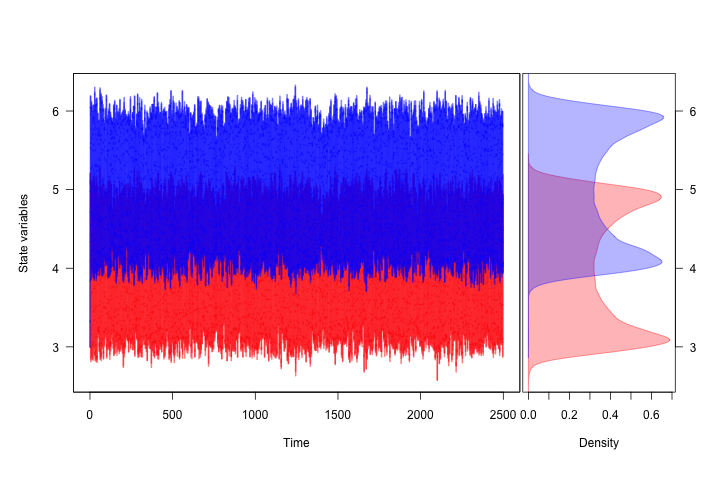}
\end{subfigure}\\
\begin{subfigure}[b]{0.4\textwidth}
\caption{}
\includegraphics[width=300px]{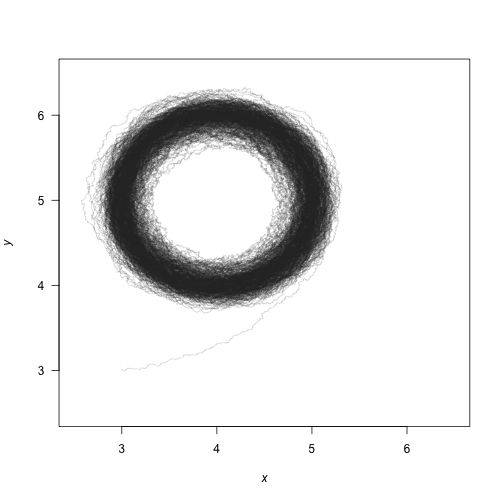}
\end{subfigure}
\caption{\textbf{(a)} A realization of system~\eqref{ex2} created using \code{TSPlot()}, with $x$ in blue and $y$ in red. The left side of (a) shows the time series. The right side of (a), which is enabled with the default \code{dens = TRUE}, shows a histogram of the $x$ and $y$ values over the entire realization. \textbf{(b)} The realization plotted in $(x,y)$-space (\code{dim = 2} in the function \code{TSPlot()}).}
\label{fig:ex2ts}
\end{figure}

\begin{figure}[ht]
\centering
\includegraphics[width=300px]{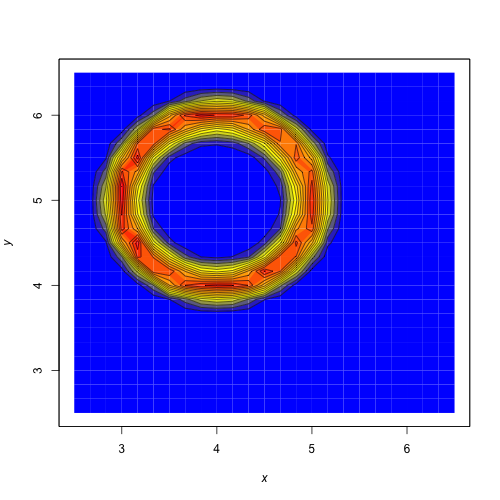}
\caption{A density plot obtained from a realization of system~\eqref{ex2} using \code{TSDensity()} with \code{dim = 2}. Red corresponds to high density, and blue to low density.}
\label{fig:ex2dens}
\end{figure}

Figure~\ref{fig:ex2ts} shows a time series for a realization of~\eqref{ex2} with $\sigma=0.1$,
$\Delta t=5\times10^{-3}$, $T=250$ and initial condition $(x_{0},\, y_{0})=(3,\,3).$
Figure~\ref{fig:ex2dens} shows a density plot of a realization with the same parameters,
except $T=2.5\times10^{3}.$

\begin{example}[commandchars=\\\{\}]
var.eqn.x <- "-(y-beta) + mu*(x-alpha)*(1-(x-alpha)^2-(y-beta)^2)"
var.eqn.y <- "(x-alpha) + mu*(y-beta)*(1-(x-alpha)^2-(y-beta)^2)"
model.state <- c(x = 3, y = 3)
model.parms <- c(alpha = 4, beta = 5, mu = 0.2)
model.sigma <- 0.1
model.time <- 1000 # we used 2500 in the figures
model.deltat <- 0.005
ts.ex2 <- TSTraj(y0 = model.state, time = model.time, deltat = model.deltat, 
     x.rhs = var.eqn.x, y.rhs = var.eqn.y, parms = model.parms, sigma = model.sigma)

TSPlot(ts.ex2, deltat = model.deltat)		# Figure \ref{fig:ex2ts}a
TSPlot(ts.ex2, deltat = model.deltat, dim = 2, line.alpha = 25)	# Figure \ref{fig:ex2ts}b 
TSDensity(ts.ex2, dim = 1)	# Histogram
TSDensity(ts.ex2, dim = 2)	# Figure \ref{fig:ex2dens}
\end{example}

\subsection*{Step 3: Local quasi-potential calculation}
In this example, there are no stable equilibrium points. There is
one stable limit cycle, and this can be used to obtain a local quasi-potential.
Using $\mathbf{z}$ as the initial point for the ordered-upwind algorithm and $Lx_{1}=-0.5,$ $Ly_{1}=-0.5,$ $Lx_{2}=7.5,$ $Ly_{2}=7.5,$
$Nx=4000,$ and $Ny=4000,$ one obtains a local quasi-potential, $\Phi_{\mathbf{z}}(x,y).$ This generates the local quasi-potential
$ \Phi_{\mathbf{z}}(x,y)$.

\begin{example}
eqn.x <- "-(y-5) + (0.2)*(x-4)*(1-(x-4)^2-(y-5)^2) "
eqn.y <- "(x-4) + (0.2)*(y-5)*(1-(x-4)^2-(y-5)^2)"
eq1.qp <- QPotential(x.rhs = eqn.x, x.start = 4.15611, x.bound = c(-0.5, 7.5), 
     x.num.steps = 4000, y.rhs = eqn.y, y.start = 5.98774, 
     y.bound = c(-0.5, 7.5), y.num.steps = 4000)
\end{example}

\subsection*{Step 4: Global quasi-potential calculation}
There is only one local quasi-potential in this example, so it is the global quasi-potential,
$\Phi(x,y)=\Phi_{\mathbf{z}}(x,y).$

\subsection*{Step 5: Global quasi-potential visualization}
\begin{figure}[ht]
  \centering
  \includegraphics[width=300px]{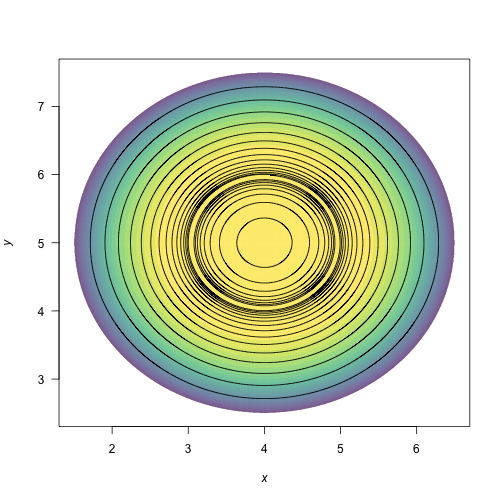}
  \caption{A contour plot of the of the quasi-potential of system~\eqref{ex2} using \code{QPContour()}. Yellow corresponds to low values of the quasi-potential, and purple to high values.}
  \label{fig:ex2qp}
\end{figure}

Figure~\ref{fig:ex2qp} shows a contour plot of the global quasi-potential.
\begin{example}
QPContour(eq1.qp, dens = c(1000, 1000), x.bound = c(-0.5, 7.5), 
	y.bound = c(-0.5, 7.5), c.parm = 10)
\end{example}

\section*{Example 3: More complicated local quasi-potential pasting}

In example 1, the procedure for pasting local quasi-potentials together into global quasi-potential was a simple, two-step process. First, one of the local quasi-potentials was translated so that the two surfaces agreed at the saddle point separating the two basins of attraction. Second, the global quasi-potential was obtained by taking the minimum of the two surfaces at each point. A general algorithm for pasting local quasi-potentials, as explained in \citet{Graham:1986cq} and \citet{Roy:1995tx}, is slightly more complicated. This process is automated in \code{QPGlobal}, but it is worth understanding the process in order to correctly interpret the outputs.

To understand the full algorithm, consider the following model:

\begin{flalign}
\label{ex3}
dX=X\left(\left(1+\alpha_{1}\right)-X^{2}-X\, Y-Y^{2}\right)\, dt+\sigma\, dW_{1} \nonumber \\
dY=Y\left(\left(1+\alpha_{2}\right)-X^{2}-X\, Y-Y^{2}\right)\, dt+\sigma\, dW_{2}.
\end{flalign}
For this analysis, let $\alpha_{1}=1.25$ and $\alpha_{2}=2.$

\subsection*{Step 1: Analyzing the deterministic skeleton}
\begin{figure}[ht]
  \centering
  \includegraphics[width=300px]{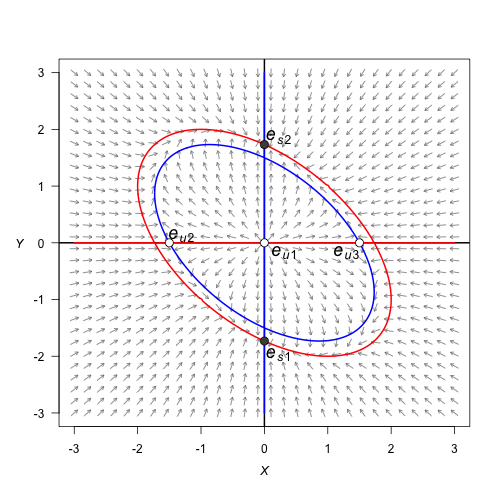}
  \caption{A stream plot of the deterministic skeleton of system~\eqref{ex3}. The blue line is an $x$-nullcline (where $\frac{dx}{dt}=0$) and the red line is a $y$-nullcline (where $\frac{dy}{dt}=0$). Open circles are stable equilibria and filled circles are unstable equilibria. Made using the package \pkg{phaseR}.}
  \label{fig:ex3stream}
\end{figure}

The deterministic skeleton of this system has five equilibria. These
are $\mathbf{e}_{u1}=(0,0)$, $\mathbf{e}_{s1}=(0,\,-1.73205),$ $\mathbf{e}_{s2}=(0,\,1.73205),$
$\mathbf{e}_{u2}=(-1.5,\,0)$ and $\mathbf{e}_{u3}=(1.5,\,0).$ The
eigenvalue analysis shows that $\mathbf{e}_{u1}$ is an unstable node,
$\mathbf{e}_{s1}$ and $\mathbf{e}_{s2}$ are stable nodes, $\mathbf{e}_{u2}$
and $\mathbf{e}_{u3}$ are saddles. Figure \ref{fig:ex3stream} shows a stream
plot of the deterministic skeleton of~\eqref{ex3}. The basin of attraction
for $\mathbf{e}_{s1}$ is the lower half-plane, and the basin of attraction
for $\mathbf{e}_{s2}$ is the upper half-plane.

\subsection*{Step 2: Stochastic simulation}
\begin{figure}[ht]
\begin{subfigure}[b]{0.4\textwidth}
\caption{}
\includegraphics[width=400px]{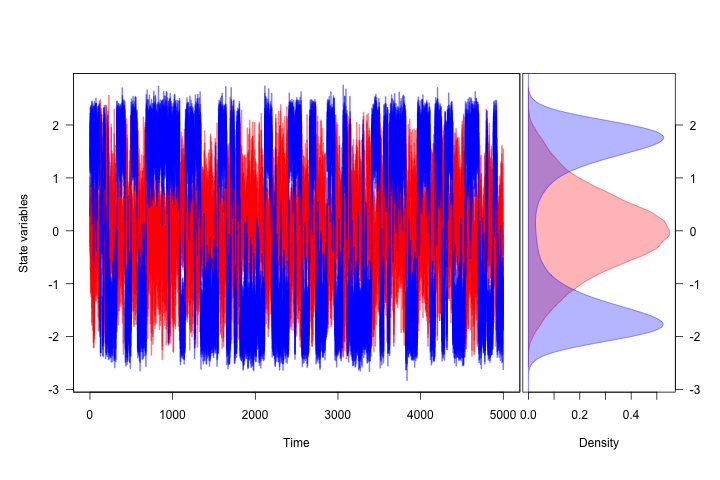}
\end{subfigure}\\
\begin{subfigure}[b]{0.4\textwidth}
\caption{}
\includegraphics[width=300px]{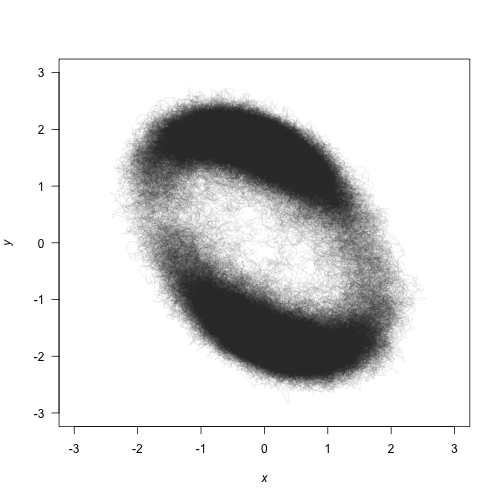}
\end{subfigure}
\caption{\textbf{(a)} A realization of system~\eqref{ex3} created using \code{TSPlot()}, with $x$ in blue and $y$ in red. The left side in panel (a) shows the time series. The right side in panel (a), which is enabled by default with parameter \code{dens  = T} in the function \code{TSPlot()}, shows a histogram of the $x$ and $y$ values over the entire realization. \textbf{(b)} The realization plotted in $(x,y)$-space with \code{TSPlot()} with \code{dim = 2}.}
\label{fig:ex3ts} 
\end{figure}

\begin{figure}[ht]
\centering
\includegraphics[width=300px]{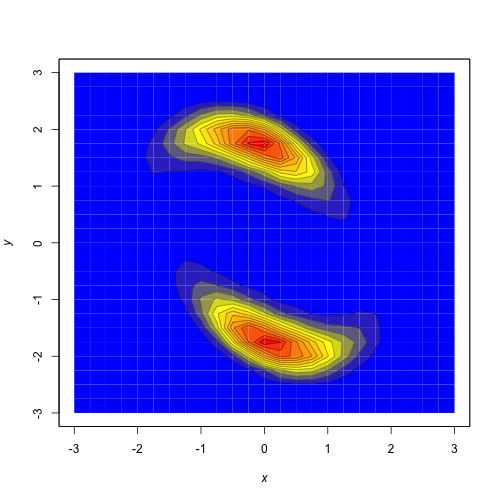}
\caption{A density plot obtained from a realization of system \eqref{ex3} by using the function \code{TSDensity()} with \code{dim = 2}, \code{contour.levels = 20}, and \code{contour.lwd = 0.1}. Red corresponds to high density, and blue to low density. }
\label{fig:ex3dens}
\end{figure}

Figure~\ref{fig:ex3ts} shows a time series for a realization of system~\eqref{ex3} with $\sigma=0.8$,
$\Delta t=0.01$, $T=5000$ and initial condition $(x_{0},\, y_{0})=(0.5,\,0.5).$
Figure~\ref{fig:ex3dens} shows a density plot of this realization.

\begin{example}[commandchars=\\\{\}]
var.eqn.x <- "x*((1+alpha1)-x*x-x*y-y*y)"
var.eqn.y <- "y*((1+alpha2)-x*x-x*y-y*y)"
model.state <- c(x = 0.5, y = 0.5)
model.parms <- c(alpha1 = 1.25, alpha2 = 2)
model.sigma <- 0.8
model.time <- 5000
model.deltat <- 0.01
ts.ex3 <- TSTraj(y0 = model.state, time = model.time, deltat = model.deltat, 
     x.rhs = var.eqn.x, y.rhs = var.eqn.y, parms = model.parms, sigma = model.sigma)

TSPlot(ts.ex3, deltat = model.deltat)				# figure \ref{fig:ex3ts}a
TSPlot(ts.ex3, deltat = model.deltat, dim = 2 , line.alpha = 25)	# figure \ref{fig:ex3ts}b
TSDensity(ts.ex3, dim = 1)				# Histogram of time series
TSDensity(ts.ex3, dim = 2 , contour.levels = 20 , contour.lwd = 0.1) # figure \ref{fig:ex3dens}
\end{example}

\subsection*{Step 3: Local quasi-potential calculation}
Two local quasi-potentials need to be calculated, $\Phi_{1}(x,y)$
corresponding to $\mathbf{e}_{s1}$, and $\Phi_{2}(x,y)$ corresponding
to $\mathbf{e}_{s2}$. In both cases, sensible boundary and mesh choices are $Lx_{1}=-3,$ $Ly_{1}=-3,$
$Lx_{2}=3,$ $Ly_{2}=3,$ $Nx=6000,$ and $Ny=6000.$

\begin{example}
equation.x = "x*((1+1.25)-x*x-x*y-y*y)"
equation.y = "y*((1+2)-x*x-x*y-y*y)"
bounds.x = c(-3, 3)
bounds.y = c(-3, 3)
step.number.x = 6000
step.number.y = 6000
eq1.x = 0
eq1.y = -1.73205
eq2.x = 0
eq2.y = 1.73205

eq1.local <- QPotential(x.rhs = equation.x, x.start = eq1.x, x.bound = bounds.x, 
     x.num.steps = step.number.x, y.rhs = equation.y, y.start = eq1.y, y.bound = 
     bounds.y, y.num.steps = step.number.y)
eq2.local <- QPotential(x.rhs = equation.x, x.start = eq2.x, x.bound = bounds.x, 
     x.num.steps = step.number.x, y.rhs = equation.y, y.start = eq2.y, y.bound = 
     bounds.y, y.num.steps = step.number.y)
\end{example}

\subsection*{Step 4: Global quasi-potential}
If one were to naively try to match the local quasi-potentials at $\mathbf{e}_{u2}$, then they would not match at $\mathbf{e}_{u3}$, and vice versa. To overcome this problem, it is necessary to think more carefully about how trajectories transition between basins of attraction. This issue can be dealt with rigorously \citep{Graham:1986cq, Roy:1995tx}, but the general principles are outlined here. Let $\Omega_{1}$ be the basin of attraction corresponding to $\mathbf{e}_{s1}$ and $\Omega_{2}$ be the basin of attraction corresponding to $\mathbf{e}_{s2}$. Let $\partial \Omega$ be the separatrix between these two basins (i.e., the $x$-axis). The most probable way for a trajectory to transition from $\Omega_{1}$ to $\Omega_{2}$ involves passing through the lowest point on the surface specified by $\Phi_{1}$ along $\partial \Omega$. Examination of $\Phi_{1}$ indicates that this point is $\mathbf{e}_{u2}$. In the small-noise limit, the transition rate from $\Omega_{1}$ to $\Omega_{2}$ will correspond to $\Phi_{1}\left(\mathbf{e}_{u2}\right)$. Similarly, the transition rate from $\Omega_{2}$ to $\Omega_{1}$ will correspond to $\Phi_{2}\left(\mathbf{e}_{u3}\right)$. The transition rate into $\Omega_{1}$ must equal the transition rate out of $\Omega_{2}$. Therefore, the two local quasi-potentials should be translated so that the minimum heights along the separatrix are the same. In other words, one must define translated local quasi-potentials $\Phi^{*}_{1}(x,y)=\Phi_{1}(x,y)+c_{1}$ and $\Phi^{*}_{2}(x,y)=\Phi_{2}(x,y)+c_{2}$ so that
\[\min{\left(\Phi^{*}_{1}(x,y)\vert(x,y)\in\partial\Omega\right)}=\min{\left(\Phi^{*}_{2}(x,y)\vert(x,y)\in\partial\Omega\right)}.
\]
In example 1, the minima of both local quasi-potentials occurred at the same point, so the algorithm amounted to matching at that point. In example 3, the minimum saddle for $\Phi_{1}$ is $\mathbf{e}_{u2}$ and the minimum saddle for $\Phi_{2}$ is $\mathbf{e}_{u3}$; the heights of the surfaces at these respective points should be matched. Thus, $c_1 = \Phi_2(e_{u3}) - \Phi_1(e_{u3})$ and $c_2 = \Phi_1(e_{u2}) - \Phi_2(e_{u2})$.  Conveniently in example 3, this is satisfied without requiring any translation (one can use $c_{1}=c_{2}=0$). Finally, the global quasi-potential is found by taking the minimum value of the matched local quasi-potentials at each point. This process is automated in \pkg{QPot}, but users can also manipulate the local quasi-potential matrices manually to verify the results. This is recommended when dealing with unusual or complicated separatrices. The code below applies the automated global quasi-potential calculation to example 3. 
\begin{example}
ex3.global <- QPGlobal(local.surfaces = list(eq1.local, eq2.local),unstable.eq.x 
     = c(0, -1.5, 1.5), unstable.eq.y = c(0, 0, 0), x.bound = bounds.x, y.bound = 
     bounds.y)
\end{example}

\subsection*{Step 5: Global quasi-potential visualization}
\begin{figure}[ht]
  \centering
  \includegraphics[width=300px]{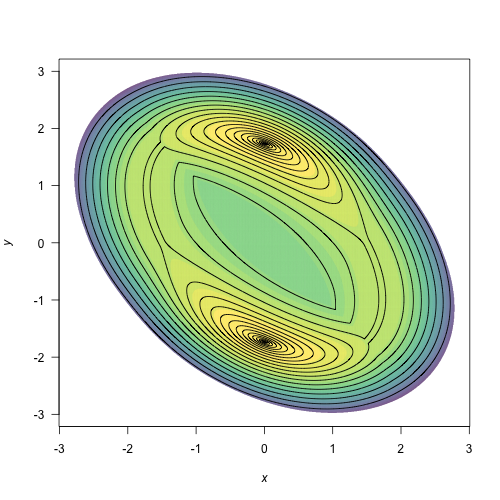}
  \caption{A contour plot of the quasi-potential of system \eqref{ex3} using the function \code{QPContour()}. Yellow corresponds to low values of the quasi-potential, and purple to high values. }
  \label{fig:ex3qp}
\end{figure}

Figure~\ref{fig:ex3qp} shows a contour plot of the global quasi-potential. Note that the surface is continuous, but not smooth. The lack of smoothness is a generic feature of global quasi-potentials created from pasting local quasi-potentials. Cusps usually form when switching from the part of solution obtained from one local quasi-potential to the other.

\begin{example}
QPContour(ex3.global, dens = c(1000, 1000), x.bound = bounds.x, y.bound = 
     bounds.y, c.parm = 5)
\end{example}

\section*{Boundary behavior}
It is important to consider the type of behavior that should
be enforced at the boundaries and on coordinate axes ($x=0$ and $y=0$). By default, the ordered-upwind method computes the quasi-potential for the system defined by the user, without regard for the influence of the boundaries or the significance of these axes.
In some cases, however, a model is only valid in a subregion of phase
space.  For example, in many population models, only the non-negative
phase space is physically meaningful. In such cases, it is undesirable
to allow the ordered-upwind method to consider trajectories that pass through negative phase space. 
In the default mode for \code{QPotential()}, if $(x,y)$ lies in positive phase space, $\Phi(x,y)$ can be impacted by the vector field in negative phase space, if the path corresponding to the minimum ``work'' passes through negative phase space. The argument \code{bounce = 'd'} corresponds to this (d)efault behavior. A user can prevent the ordered upwind method from passing trajectories through negative phase space by using the option \code{bounce = 'p'} for (p)ositive values only. This option can be interpreted as a reflecting boundary condition. It forces the front of solutions obtained by the ordered upwind method to stay in the defined boundaries, in this case the positive phase space. A more generic option is \code{bounce = 'b'} for (b)ounce, which reflects based on the user-supplied boundaries. Even using this option, it is still wise to have padding space along coordinate boundaries to prevent premature termination of the algorithm, which is set with \code{bounce.edge}. 

\section*{Different noise terms}
In the cases considered so far, the noise terms for the $X$ and $Y$
variables have had identical intensity. This was useful for purposes
of illustration in the algorithm, but it will often be untrue of real-world
systems. Fortunately, \pkg{QPot} can accomodate other noise terms with coordinate transforms. Consider a system
of the form:
\begin{flalign}
\label{eqsdnt1}
dX=f_{1}(X,Y)\, dt+\sigma\, g_{1}\, dW_{1} \nonumber\\
dY=f_{2}(X,Y)\, dt+\sigma\, g_{2}\, dW_{2}.
\end{flalign}
$\sigma$ is a scaling parameter that specifies the
overall noise intensity. The parameters $g_{1}$ and $g_{2}$ specify
the relative intensity of the two noise terms. To transform this system
into a form that is useable for \pkg{QPot}, make the change of variable
$\tilde{X}=g_{1}^{-1}X$ and $\tilde{Y}=g_{2}^{-1}Y.$ In the new
coordinates, the drift terms (that is, the terms multiplied by $dt$),
will be different. These new drift terms can be incorporated into the deterministic skeleton that is input into \pkg{QPot}. After obtaining the global quasi-potential for these transformed coordinates, one can switch back to the original
coordinates for plotting.

Many models contain multiplicative noise terms. These are of the form:
\begin{flalign}
\label{eqsdnt1}
dX=f_{1}(X,Y)\, dt+\sigma\, g_{1}\, X\, dW_{1} \nonumber \\
dY=f_{2}(X,Y)\, dt+\sigma\, g_{2}\, Y\, dW_{2}.
\end{flalign}
To transform this system into a form that is useable for \pkg{QPot}, make
the change of variable $\tilde{X}=g_{1}^{-1}\ln\left(X\right)$ and
$\tilde{Y}=g_{1}^{-1}\ln\left(Y\right).$ This coordinate change is
non-linear, so It\^{o}'s lemma introduces extra terms into the drift of the
transformed equations. If $\sigma$ is small, though,
these terms can be discounted, and the new drift terms will
remain independent of $\sigma.$ These new drift terms can be input into \pkg{QPot}. After obtaining the global quasi-potential for these transformed coordinates, one can switch back to the original
coordinates.

\section*{Conclusion}
\pkg{QPot} is an R package that provides several important tools for analyzing two-dimensional systems of stochastic differential equations. These include functions for generating realizations of the stochastic differential equations, and for analyzing and visualizing the results. A central component of \pkg{QPot} is the calculation of quasi-potential functions, which are highly useful for studying stochastic dynamics. For example, quasi-potential functions can be used to compare the stability of different attractors in stochastic systems, a task that traditional linear stability analysis is poorly suited for \citep{nolting2015}. By offering an intuitive way to quantify attractor stability, quasi-potentials are poised to become an important means of understanding phenomena like metastability and alternative stable states. \pkg{QPot} makes quasi-potentials accessible to R users interested in applying this new framework.

\section*{Author contributions}
K.C.A, C.M.M., B.C.N., and C.R.S. designed the project.  M.K.C. wrote the C code for finding the quasi-potential;  C.M.M. and C.R.S. wrote the R code and adapted the C code.

\section*{Acknowledgements}
This work was supported by a Complex Systems Scholar grant to K.C.A. from the James S. McDonnell Foundation.  M.K.C. was partially supported by NSF grant 1217118.

\bibliography{RJournalRefs.bib}

\begin{thebibliography}{17}
\providecommand{\natexlab}[1]{#1}
\providecommand{\url}[1]{\texttt{#1}}
\expandafter\ifx\csname urlstyle\endcsname\relax
  \providecommand{\doi}[1]{doi: #1}\else
  \providecommand{\doi}{doi: \begingroup \urlstyle{rm}\Url}\fi

\bibitem[Adler et~al.(2015)Adler, Murdoch, Nenadic, Urbanek, Chen, Gebhardt,
  Bolker, Csardi, Strzelecki, and Senger]{adler:2015aa}
D.~Adler, D.~Murdoch, O.~Nenadic, S.~Urbanek, M.~Chen, A.~Gebhardt, B.~Bolker,
  G.~Csardi, A.~Strzelecki, and A.~Senger.
\newblock rgl: 3d visualization device system for {R} using {O}pen{GL}, 2015.
\newblock URL \url{https://cran.r-project.org/package=rgl}.

\bibitem[Allen(2007)]{Allen:2007ww}
E.~J. Allen.
\newblock \emph{{Modeling with Ito stochastic differetnial equations}},
  volume~22 of \emph{Mathematical modelling: theory and applications}.
\newblock Springer, 2007.

\bibitem[Cameron(2012)]{Cameron:2012ex}
M.~K. Cameron.
\newblock {Finding the quasipotential for nongradient SDEs}.
\newblock \emph{Physica D}, 241\penalty0 (18):\penalty0 1532--1550, 2012.

\bibitem[Chua and Parker(1989)]{chua}
T.~S. P.~L. Chua and T.~S. Parker.
\newblock \emph{Practical umerical algorithms for chaotic systems}, chapter~5.
\newblock Springer-Verlag, 1989.

\bibitem[Collie and Spencer(1994)]{collie:1994aa}
J.~S. Collie and P.~D. Spencer.
\newblock Modeling predator-prey dynamics in a fluctuating environment.
\newblock \emph{Canadian Journal of Fisheries and Aquatic Sciences},
  51\penalty0 (12):\penalty0 2665--2672, 1994.

\bibitem[Freidlin and Wentzell(2012)]{Freidlin:2012wd}
M.~I. Freidlin and A.~D. Wentzell.
\newblock \emph{{Random perturbations of dynamical systems}}, volume 260.
\newblock Springer, 2012.

\bibitem[Graham and T{\'e}l(1986)]{Graham:1986cq}
R.~Graham and T.~T{\'e}l.
\newblock {Nonequilibrium potential for coexisting attractors}.
\newblock \emph{Physical Review. A.}, 33\penalty0 (2):\penalty0 1322--1337,
  1986.

\bibitem[Grayling(2014)]{grayling:2014aa}
M.~J. Grayling.
\newblock phase{R}: An {R} package for phase plane analysis of autonomous ode
  systems.
\newblock \emph{The {R} Journal}, pages 43--51, 2014.

\bibitem[Iacus(2009)]{iacus:2009aa}
S.~M. Iacus.
\newblock \emph{Simulation and inference for stochastic differential equations:
  with R examples}, volume~1.
\newblock Springer Science \& Business Media, 2009.

\bibitem[Nolting and Abbott(Accepted)]{nolting2015}
B.~C. Nolting and K.~C. Abbott.
\newblock Balls, cups, and quasi-potentials: quantifying stability in
  stochastic systems.
\newblock \emph{Ecology}, Accepted.

\bibitem[Roy and Nauman(1995)]{Roy:1995tx}
R.~V. Roy and E.~Nauman.
\newblock {Noise-induced effects on a non-linear oscillator}.
\newblock \emph{Journal of Sound and Vibration}, 1995.

\bibitem[Sethian and Vladimirsky(2001)]{sethian:2001aa}
J.~A. Sethian and A.~Vladimirsky.
\newblock Ordered upwind methods for static {H}amilton-{J}acobi equations.
\newblock \emph{Proceedings of the National Academy of Sciences}, 98\penalty0
  (20):\penalty0 11069--11074, 2001.
\newblock \doi{10.1073/pnas.201222998}.

\bibitem[Sethian and Vladimirsky(2003)]{sethian:2003aa}
J.~A. Sethian and A.~Vladimirsky.
\newblock Ordered upwind methods for static {H}amilton-{J}acobi equations:
  Theory and algorithms.
\newblock \emph{SIAM Journal on Numerical Analysis}, 41\penalty0 (1):\penalty0
  325--363, 2003.
\newblock \doi{10.1137/S0036142901392742}.

\bibitem[Soetaert(2013)]{soetaert:2013aa}
K.~Soetaert.
\newblock plot3d: Plotting multi-dimentional data in {R}, 2013.
\newblock URL \url{https://cran.r-project.org/package=plot3D}.

\bibitem[Soetaert and Herman(2008)]{soetaert:2008aa}
K.~Soetaert and P.~M. Herman.
\newblock \emph{A practical guide to ecological modelling: using {R} as a
  simulation platform}.
\newblock Springer Science \& Business Media, 2008.

\bibitem[Soetaert et~al.(2010)Soetaert, Petzoldt, and Setzer]{soetaert:2010aa}
K.~Soetaert, T.~Petzoldt, and R.~W. Setzer.
\newblock Solving differential equations in {R}: Package de{S}olve.
\newblock \emph{Journal of Statistical Software}, 33\penalty0 (9):\penalty0
  1--25, 2010.

\bibitem[Steele and Henderson(1981)]{steele:1981aa}
J.~Steele and E.~Henderson.
\newblock A simple plankton model.
\newblock \emph{American Naturalist}, pages 676--691, 1981.

\end{thebibliography}


\address{Christopher M. Moore\\
  Department of Biology\\
  Case Western Reserve University\\
  United States\\}
\email{life.dispersing@gmail.com}

\address{Christopher R. Stieha\\
  Department of Biology\\
  Case Western Reserve University\\
  United States\\}
\email{stieha@hotmail.com}

\address{Ben C. Nolting\\
  Department of Biology\\
  Case Western Reserve University\\
  United States\\}
\email{ben.nolting@case.edu}

\address{Maria K. Cameron\\
  Department of Mathematics\\
  University of Maryland\\
  United States\\}
\email{cameron@math.umd.edu}

\address{Karen C. Abbott\\
  Department of Biology\\
  Case Western Reserve University\\
  United States\\}
\email{kcabbott@case.edu}

\section{Figures}

\end{article}
\end{document}